\documentclass[10pt,conference,letterpaper]{IEEEtran}

\usepackage{cite}
\usepackage{epsfig}
\usepackage{times}
\usepackage{amsmath}
\usepackage{amssymb}
\usepackage{psfrag}

\usepackage[latin1]{inputenc}
\usepackage{graphicx}
\usepackage{latexsym}

\newtheorem{theorem}{\it Theorem}
\newtheorem{lemma}{\it Lemma}

\newtheorem{example}{\it Example}
\newtheorem{definition}{\it Definition}

\begin{document}
\title{The Capacity of Random Linear Coding Networks as Subspace Channels}



\author{\authorblockN{Bartolomeu F. Uchôa-Filho and Roberto W. Nóbrega}
\authorblockA{Communications Research Group \\ Department of Electrical Engineering \\ Federal University
of Santa Catarina \\ Florianópolis, SC, 88040-900, Brazil \\ E-mail: { \{uchoa,rwnobrega\}@eel.ufsc.br}}}

\maketitle

\begin{abstract}
In this paper, we consider noncoherent random linear coding networks (RLCNs) as a discrete memoryless channel (DMC) whose input and output alphabets consist of subspaces. This contrasts with previous channel models in the literature which assume matrices as the channel input and output. No particular assumptions are made on the network topology or the transfer matrix, except that the latter may be rank-deficient according to some rank deficiency probability distribution. We introduce a random vector basis selection procedure which renders the DMC symmetric. The capacity we derive can be seen as a lower bound on the capacity of noncoherent RLCNs, where subspace coding suffices to achieve this bound.
\end{abstract}


\section{Introduction}
\label{sec1}

Random linear network coding (RLNC), proposed by Ho {\em et al.}~\cite{ho.03}, is a very attractive way to obtain the benefits of linear network coding~\cite{Ahlswede.00,LNC.03} in dynamically changing or large scale networks. In a decentralized way, by requiring the intermediate nodes to randomly combine incoming packets over a finite field, RLNC achieves the network min-cut with probability approaching one as the field size goes to infinity. For this reason, RLNC is an important research topic and has recently received a great deal of attention. Following~\cite{Yang.09}, we call a network employing RLNC a {\it random linear coding network} (RLCN).

In order to recover the transmitted packets, a receiving node needs to know its associated network transfer matrix. In reality, the transfer matrix depends on the network topology and the network code used. However, to allow for some analysis, in previous works on RLCNs authors have made different assumptions on the characteristics of the transfer matrix, which has been assumed to have full rank~\cite{Silvacapacity.08, Monta.07}, bounded rank deficiency~\cite{Silva.08}, or entries selected uniformly at random~\cite{Jafari.08,Jafari.09}. Also, it has been considered as being selected either by an adversary~\cite{Silva.08} or uniformly at random among all nonsingular matrices~\cite{Silvacapacity.08}.

In some works, the approach for the capacity of RLCNs is combinatorial. For example, in~\cite{KS.08}, where the authors consider the noncoherent scenario in which the transfer matrix is assumed unknown to both the transmitter and the receiver, a {\em subspace coding} approach is prescribed and coding size bounds under a topology-independent operator channel model are provided.

A probabilistic approach for the capacity of RLCNs was first presented by Montanari and Urbanke~\cite{Monta.07}, who have considered an additive perturbation (error) matrix chosen uniformly at random from all matrices of given dimension and rank. Later, their work was further extended by Silva {\it et al.}~\cite{Silvacapacity.08}, who have considered both the transfer and the error matrices as chosen according to more general probability distributions.

Recently, Yang and Yang~\cite{Yang.09} have called the attention to the importance of understanding how network topologies affect RLNC. More specifically, they studied a particular family of unicast networks and identified a worst-case unicast network topology: the S-networks. They derived a capacity formula for S-networks, and showed that for this topology subspace coding is optimal in the sense that it suffices to achieve capacity. Similar conclusions regarding the use of subspace coding were reached in~\cite{Jafari.08, Jafari.09}. However, in these works, RLCNs are seen as a channel not with subspaces but with matrices as inputs and outputs.

In this paper, we consider the noncoherent scenario of~\cite{KS.08} and make no particular assumption on the network topology or the network code used. Since the transmission strategy using subspaces is oblivious to the underlying network topology and to the particular linear combinations performed at the network nodes~\cite{KS.08}, and since no other coding approach seems more suitable to the noncoherent scenario, we begin by treating RLCNs as a {\em discrete memoryless channel} (DMC) with subspaces as inputs and outputs. We build up a channel model for RLCNs that depends solely on the dimension of the subspaces that constitute the input alphabet and the rank deficiency probability distribution associated with the RLCN. The problem is that it is difficult to characterize such channel because, as we will see, under certain conditions, the received subspace depends on the choice of the transmitted matrix. Also, due to the many forms the transfer matrix can take, it is difficult, if not impossible, to characterize this channel simply from the rank deficiency probability distribution.

We overcome these drawbacks by introducing a random basis selector, a device which selects a matrix uniformly at random from the set of all ordered bases for the subspace. This idea has been mentioned in~\cite{KS.08} as a way to saying that in subspace coding there is no benefit in choosing a particular basis for transmitting a given subspace. A similar randomization procedure is also followed in~\cite{Monta.07}, with the purpose of facilitating the decoding algorithm. Completely innocuous from the subspace coding perspective, this random basis selection plays a fundamental role in the characterization of the channel and in the derivation of its capacity, as we show in the present paper.

Interestingly, the capacity formula we derive exactly corresponds to the one in~\cite{Yang.09}, which has been derived for a specific network topology. Our conclusions regarding noncoherent RLCNs parallel those in~\cite{Yang.09}: while Yang and Yang conclude that the capacity of unicast networks with given topology is lower bounded by the capacity of S-networks, we conclude that the same lower bound applies to the capacity of noncoherent RLCNs for general (unknown) topology, and that subspace coding is sufficient to achieve this bound.

The paper is organized as follows. In the next section, we describe the system model and review some important concepts for the forthcoming derivations. In Section \ref{sec3}, where the main contribution of this work is presented, we characterize the subspace channel model for RLCNs and derive its channel capacity. Finally, in Section \ref{sec4}, we present our conclusions and final comments.

\section{Preliminaries and Notation}
\label{sec2}

In this paper, random entities are represented using boldface letters, while italic letters are used for their samples. However, the letters $O$, $U$, and $V$ are reserved to denote subspaces.

\subsection{Random Linear Coding Networks as a Matrix Channel}

Consider a time-varying network with one source node and one sink node (unicast), with the intermediate nodes performing uniform random linear combining over a finite field $\mathrm{I\!F}_q$, where $q$ is a prime power. At time slot $t$, the matrix channel for this network is given by
\begin{equation}
\mathbf{Y}(t)=\mathbf{G}(t)\mathbf{X}(t),
\label{eq_Y}
\end{equation}
where $\mathbf{X}(t)$ is the $m \times T$ transmitted matrix and $\mathbf{Y}(t)$ is the $n \times T$ received matrix. The rows of $\mathbf{X}(t)$ (resp., $\mathbf{Y}(t)$) represent the $m$ (resp. $n$) transmitted (resp., received) packets, which are seen as vectors over $\mathrm{I\!F}_q$. $\mathbf{G}(t)$ is the network transfer matrix, an $n \times m$ random matrix over $\mathrm{I\!F}_q$ whose entries depend on the network topology and the network code used. The time slot is the time during which $\mathbf{G}(t)$ remains constant, and also corresponds to $T$ uses of the network, where $T$ is the packet size. In this paper, we assume that both the transfer matrix and the network topology are unknown to both the transmitter and the receiver. Following~\cite{Silvacapacity.08, Monta.07}, however, we assume for simplicity a square transfer matrix with known dimension $m=n=h$ fixed at all times.

Due to unfortunate selection of random coefficients, the network transfer matrix may be rank-deficient. According to the channel model in (\ref{eq_Y}),  the rank deficiency of $\mathbf{G}(t)$, herein denoted as ${\overline{\tau}}_{\mathbf{G}}(t)$, is a discrete-time stochastic process. We assume that ${\overline{\tau}}_{\mathbf{G}}(t)$ is first-order stationary so that a probability distribution of ${\overline{\tau}}_{\mathbf{G}}(t)$, which is independent of $t$, exists and is given by $p_{{\overline{\tau}}_{\mathbf{G}}}(r)$, where $r=0,\ldots,h$.

In practice, if this process is ergodic, its probability distribution can be estimated at the sink node by continuously updating a histogram based on the rank deficiency\footnote{Recall that, if $\mathbf{X}$ is full-rank then ${\overline{\tau}}_{\mathbf{Y}}={\overline{\tau}}_{\mathbf{G}}$ if $\mathbf{Y}=\mathbf{G}\mathbf{X}$.} of the received matrix $\mathbf{Y}(t)$. The probability distribution can then be communicated to the source node through a low-rate feedback channel, either from time to time or once and for all when a steady distribution is reached. From now on, we will drop the time index $t$ in all variables for convenience.

\subsection{Subspace Coding}
\label{sec_SC}

The idea behind subspace coding is that, since the received vectors are linear combinations of the transmitted vectors, the vector space spanned by the transmitted vectors is preserved. So, in subspace coding, information is encoded in the choice of a subspace of the vector space $\mathrm{I\!F}_q^T$~\cite{KS.08}. We denote by $\left\langle \mathbf{X}\right\rangle$ the subspace spanned by the rows of the transmitted matrix~$\mathbf{X}$.

A {\em subspace code} is then a subset of the {\em projective space} ${\cal P}(\mathrm{I\!F}_q^T)$, which is the collection of all subspaces of $\mathrm{I\!F}_q^T$. A codeword of a subspace code is one of these subspaces. This contrasts with the notion of classical linear block coding, where a code is a subspace and a codeword is a vector in this subspace.

Most known subspace codes are {\em constant-dimension} codes, or codes in the Grassmannian, {\it i.e.}, subspace codes whose all codewords have the same dimension. We denote the Grassmannian of dimension $h$ by ${\cal P}(\mathrm{I\!F}_q^T,h)$, so that ${\cal P}(\mathrm{I\!F}_q^T)= \bigcup_{h=0}^T {\cal P}(\mathrm{I\!F}_q^T,h)$. The number of distinct $\ell$-dimensional subspaces of an $n$-dimensional vector space over $\mathrm{I\!F}_q$ is given by
\[ \displaystyle \dbinom{n}{\ell}_{q} = \prod_{i=0}^{\ell - 1} \dfrac{q^{n-i}-1}{q^{\ell-i}-1}  , \]
known as the $q$-ary {\em Gaussian coefficient}.

Although the use of subspaces of multiple dimensions is optimal under certain circumstances~\cite{Jafari.09}, there has been an increasing interest in constant-dimension codes, chiefly due to their simpler encoding and decoding algorithms.

In~\cite{Silva.08}, Silva {\it et al.} have given an important construction of constant-dimension codes which builds upon a {\em lifting} of a class of {\em maximum rank-metric codes} due to Gabidulin~\cite{Gabidulin.85}. More recently, in~\cite{Silvaisit.09}, Silva and Kschischang have presented new fast encoding and decoding algorithms for Gabidulin codes that can be used towards their lifted counterpart.

\subsection{Discrete Memoryless Channels}
A DMC is defined by the triplet $({\cal X}, p_{\mathbf{Y}|\mathbf{X}}, {\cal Y})$, where ${\cal X}$ (resp., ${\cal Y}$) is the channel input (resp., output) alphabet, and $p_{\mathbf{Y}|\mathbf{X}}(Y|X)$ is the conditional probability that $Y \in {\cal Y}$ is received given that $X \in {\cal X}$ is sent. The channel is {\em memoryless} in the sense that what happens to the transmitted symbol at one time is independent of what happens to the transmitted symbol at any other time. Let $\mathbf{X}$ and $\mathbf{Y}$ be the random entities representing the input and output symbols, respectively. The {\em channel capacity} of the DMC is then given by
\begin{eqnarray}
\label{eq_C_DMC} \displaystyle
C & = &  \max_{p_{\mathbf{X}}(X)} I(\mathbf{X};\mathbf{Y}) \\
& = & \max_{p_{\mathbf{X}}(X)} H(\mathbf{X}) - H(\mathbf{X}|\mathbf{Y}) = \max_{p_\mathbf{X}(X)} H(\mathbf{Y}) - H(\mathbf{Y}|\mathbf{X}), \nonumber
\end{eqnarray}
where $I(\mathbf{X};\mathbf{Y})$, $H(\mathbf{X})$ (resp., $H(\mathbf{Y})$), and $H(\mathbf{X}|\mathbf{Y})$ (resp., $H(\mathbf{Y}|\mathbf{X})$) are the mutual information between $\mathbf{X}$ and $\mathbf{Y}$, the entropy of $\mathbf{X}$ (resp., $\mathbf{Y}$), and the conditional entropy of $\mathbf{Y}$ (resp., $\mathbf{X}$) given $\mathbf{X}$ (resp., $\mathbf{Y}$), respectively.
The maximization in (\ref{eq_C_DMC}) is over all possible input distributions $p_{\mathbf{X}}$.

The points of the distribution $p_{\mathbf{Y}|\mathbf{X}}(Y|X)$ for all $X \in {\cal X}$ and all $Y \in {\cal Y}$ can be arranged in a matrix, called the {\em transition probability matrix}, where the $X$'s (resp., $Y$'s) are related to the rows (resp., columns).

A DMC can be pictorially described by a bipartite graph, where the nodes at left (resp., right) represent the input (resp., output) alphabet. Each input (resp., output) node has a number of sockets from (resp., to) which edges emanate (resp., connect). If an output symbol occurs with some nonzero probability given an input symbol, then there is an edge connecting the corresponding input node to the output node. Of course, the total number of sockets on the ${\cal X}$ side equals the total number o sockets on the ${\cal Y}$ side.

The following definitions and theorems have been adapted from~\cite{Cover.06,Massey.93}.
\definition \label{def_ss} A DMC is {\em uniformly dispersive} (resp., {\em focusing}) if all the rows (resp., columns) of the transition probability matrix are permutations of each other. A DMC that is both uniformly dispersive and uniformly focusing is said to be {\em strongly symmetric}.
\theorem \label{th_ss} For a strongly symmetric DMC, the capacity is given by
\[ C = \log |{\cal Y}| - H(p), \]
where $H(p)$ denotes the entropy of the probability vector ${p}$, formed by the elements of a row of the transition probability matrix. Moreover, this capacity is achieved by the uniform input distribution.

If a channel is uniformly dispersive but not uniformly focusing, it may still have a weaker form of symmetry.
\definition A uniformly dispersive DMC is said to be {\em symmetric} (also known as {\em component-symmetric}) if, for some $L$, it can be decomposed into $L$ strongly symmetric channels with selection probabilities $q_0, \ldots, q_{L-1}$.
\theorem \label{th_s} For a symmetric DMC, the capacity is given~by
\[ C = \displaystyle \sum_{i=0}^{L-1} q_i C_i, \]
where $C_i$ is the capacity of the $i$-th component strongly symmetric channel associated with the symmetric DMC.

\section{The Capacity of the Subspace Channel as a Symmetric DMC}
\label{sec3}

We want to characterize the DMC
\[
\left( {\cal X} = {\cal P}(\mathrm{I\!F}_q^T,h), p_{\left\langle \mathbf{Y}\right\rangle | \left\langle \mathbf{X}\right\rangle}, {\cal Y} = \bigcup_{r=0}^h {\cal P}(\mathrm{I\!F}_q^T,r) \right)
\]
serving as a channel model for RLCNs essentially without any assumption on the network topology or the particular network code used. The main goal is to characterize the conditional probability $p_{\left\langle \mathbf{Y}\right\rangle | \left\langle \mathbf{X}\right\rangle}$ that the subspace $\left\langle \mathbf{Y}\right\rangle$ is received when subspace $\left\langle \mathbf{X}\right\rangle$ is sent as a function of the rank deficiency probability distribution $p_{{\overline{\tau}}_{\mathbf{G}}}(r)$. The difficulties related to the characterization of this DMC, raised in the introductory section, will be illustrated by the next two examples.

\example Let $U=\{000,010,100,110\}$ be the input subspace. There are six different basis matrices for $U$, two of which are
\[ \textsl{X}=\left[ \begin{array}{ccc} 0 & 1 & 0 \\ 1 & 0 & 0 \end{array} \right], \; \textsl{X}^{\prime}=\left[ \begin{array}{ccc} 0 & 1 & 0 \\ 1 & 1 & 0 \end{array} \right].\]
Suppose that $\textsl{G}$ has rank deficiency ${\overline{\tau}}_{\textsl{G}}=1$ and is given by
\[ \textsl{G} = \left[ \begin{array}{cc} 0 & 1 \\ 0 & 1 \end{array} \right]. \]
For the corresponding received matrices, $\textsl{Y}=\textsl{G}\textsl{X}$ and $\textsl{Y}^{\prime}=\textsl{G}\textsl{X}^{\prime}$, the associated one-dimensional output subspaces are $\left\langle \textsl{Y}\right\rangle = \{ 000,100\}$ and $\left\langle \textsl{Y}^{\prime}\right\rangle = \{ 000,110\}$, respectively.

We can see from Example 1 that, given the input subspace~$U$ and a fixed non-zero, rank-deficient transfer matrix $\textsl{G}$, the output subspace may depend on the selection of the basis for $U$.
The first difficulty resides on the fact that a different DMC results for each possible combination of bases selections for input subspaces, and there is an exponentially large number of them.

Characterizing our DMC is also difficult because the network code can affect the transmitted matrix $\textsl{X}$ (and the input subspace $\left\langle \textsl{X}\right\rangle$) in too many ways. The next example illustrates the problem.
\example Consider the same input subspace $U$ in Example 1 and $\textsl{X}^{\prime}$ as the transmitted matrix. Also, suppose again that ${\overline{\tau}}_{\textsl{G}}=1$ but with the transfer matrix given by
\[ \textsl{G}^{\prime} = \left[ \begin{array}{cc} 1 & 0 \\ 1 & 0 \end{array} \right]. \]
The received matrix, $\textsl{Y}^{\prime \prime}=\textsl{G}^\prime\textsl{X}^{\prime}$, gives rise to the output subspace $\left\langle \textsl{Y}^{\prime \prime}\right\rangle = \{ 000,010\}$.

We can see from Example 2 that, for a fixed transmitted matrix $\textsl{X}^{\prime}$, different transfer matrices with the same rank deficiency may produce different output subspaces. (Note that this problem does not exist if the distinct transfer matrices have full rank.) As a result, for a fixed rank deficiency probability distribution the DMC is not fixed, but varies according to the realization of $\mathbf{G}$. In other words, a DMC for the subspace channel cannot be defined solely from rank deficiencies even if a fixed mapping $U \rightarrow \textsl{X}$ with $\left\langle \textsl{X}\right\rangle=U$ were specified for every subspace $U$ in the Grassmannian.

We remedy these problems by introducing a random basis selector. We show that, upon using a basis matrix $\mathbf{X}$ produced by this device, the resulting DMC represents RLCNs, and is both well-defined and consistent with the subspace coding approach. Given a subspace $U \in {\cal X}$ at the input of the selector, it outputs a matrix $\mathbf{X}$ which is selected, uniformly at random, from the set of all basis matrices whose rows span the subspace~$U$. Accordingly, the conditional probability of basis matrix $\mathbf{X}$ given the subspace $\left\langle \mathbf{X}\right\rangle=U$ is given by
\begin{equation}
\label{eq_Px}
p_{\mathbf{X} | \left\langle \mathbf{X}\right\rangle } (\textsl{X} | U)  = \prod_{i=1}^h \left(q^h-q^{i-1} \right)^{-1},
\end{equation}
where the inverse of (\ref{eq_Px}) is exactly the number of ordered bases that span a subspace $U$ in ${\cal P}(\mathrm{I\!F}_q^T,h)$~\cite{KS.08}. From now on, when we refer to a DMC we mean the DMC that models RLCNs under this random basis selection.

We now prove that the distribution $p_{\left\langle \mathbf{Y}\right\rangle | \left\langle \mathbf{X}\right\rangle}$ is fully characterized by the rank deficiency probability distribution only, and is conditionally independent of the network topology and the network code used.
\begin{theorem}
\label{th_1}
For the DMC \[ ( {\cal P}(\mathrm{I\!F}_q^T,h), p_{\left\langle \mathbf{Y}\right\rangle | \left\langle \mathbf{X}\right\rangle}, \bigcup_{r=0}^h {\cal P}(\mathrm{I\!F}_q^T,r) ),\]  the transition probability distribution is given by
\[ \displaystyle p_{\left\langle \mathbf{Y}\right\rangle | \left\langle \mathbf{X}\right\rangle} (V|U) = \left \{ \begin{array}{ll} \dfrac{p_{{\overline{\tau}}_{\mathbf{G}}}(h - \dim V)}{\dbinom{h}{{{\dim} V}}_{q}}, & \mbox{if } V \subseteq U\\ 0, & \mbox{else.} \end{array} \right.
\]
\end{theorem}
\begin{proof} We begin by expanding the conditional probability distribution $p_{\left\langle \mathbf{Y}\right\rangle | \left\langle \mathbf{X}\right\rangle, {\overline{\tau}}_{\mathbf{G}}}$ as follows:
\begin{eqnarray}
 \label{eq_t1a} \lefteqn{p_{\left\langle \mathbf{Y}\right\rangle | \left\langle \mathbf{X}\right\rangle, {\overline{\tau}}_{\mathbf{G}}} (V|U,\rho) } \nonumber \\ & = & \sum_{\textsl{X}} p_{\left\langle \mathbf{Y}\right\rangle, \mathbf{X} | \left\langle \mathbf{X}\right\rangle, {\overline{\tau}}_{\mathbf{G}}} (V,\textsl{X} |U,\rho) \nonumber \\ & = & \sum_{\textsl{X}} p_{\mathbf{X} | \left\langle \mathbf{X}\right\rangle, {\overline{\tau}}_{\mathbf{G}}} (\textsl{X}|U,\rho) \cdot p_{\left\langle \mathbf{Y}\right\rangle | \mathbf{X},\left\langle \mathbf{X}\right\rangle, {\overline{\tau}}_{\mathbf{G}}} (V|\textsl{X},U,\rho) \nonumber \\ & = & \sum_{\textsl{X}:\left\langle \textsl{X} \right\rangle=U} p_{\mathbf{X} | \left\langle \mathbf{X}\right\rangle} (\textsl{X}|U) \cdot p_{\left\langle \mathbf{Y}\right\rangle | \mathbf{X}, {\overline{\tau}}_{\mathbf{G}}} (V|\textsl{X},\rho) \nonumber \\
 & = & \prod_{i=1}^h \left(q^h-q^{i-1} \right)^{-1} \sum_{\textsl{X}:\left\langle \textsl{X} \right\rangle=U} p_{\left\langle \mathbf{Y}\right\rangle | \mathbf{X}, {\overline{\tau}}_{\mathbf{G}}} (V|\textsl{X}, \rho) ,
\end{eqnarray}
where we used the fact that $\mathbf{X}$ is conditionally independent of ${\overline{\tau}}_{\mathbf{G}}$ given $\left\langle \mathbf{X}\right\rangle$ and the result in (\ref{eq_Px}) for $p_{\mathbf{X} | \left\langle \mathbf{X}\right\rangle }$. The summation in (\ref{eq_t1a}) can be obtained as follows. Suppose that $\mathbf{G}$ is some fixed (unknown) matrix $\textsl{G}$ over $\mathrm{I\!F}_q$ with ${\overline{\tau}}_{\textsl{G}}=\rho = h-\dim V$. Then the transmitted matrix $\textsl{X}$ is mapped into some output subspace $\left\langle \textsl{GX}\right\rangle = V^{\prime}$ with probability one. So, the summand in (\ref{eq_t1a}) is either 1 (if $V^{\prime}=V$) or zero (if $V^{\prime} \ne V$). The resulting sum is thus the number of ordered bases $\textsl{X}$ that span the subspace $U$ in ${\cal P}(\mathrm{I\!F}_q^T,h)$ and that produce $\left\langle \textsl{GX}\right\rangle=V$ as output subspace. Since the range of $\textsl{X}$ is the set of all ordered bases that span the subspace $U$, by symmetry, the summation in (\ref{eq_t1a}) is exactly the total number of ordered bases that span the subspace $U$ divided by the total number of subspaces of dimension $\dim V$ of $U$, which is given~by
\[ \displaystyle
\dfrac{\displaystyle \prod_{i=1}^h \left(q^h-q^{i-1} \right)}{{\dbinom{h}{\dim V}_{q}}}.
\]
With this result, (\ref{eq_t1a}) becomes
\begin{eqnarray}
 \label{eq_t1b} \displaystyle
 \lefteqn{p_{\left\langle \mathbf{Y}\right\rangle | \left\langle \mathbf{X}\right\rangle, {\overline{\tau}}_{\mathbf{G}}} (V|U,\rho)  =} \nonumber \\ & &  \left \{ \begin{array}{ll} \dfrac{1}{\dbinom{h}{{h - \rho}}_{q}}, & \mbox{if } V \subseteq U, \rho = h - \dim V   \\ 0, & \mbox{else.} \end{array} \right.
 \end{eqnarray}
We now write
\begin{eqnarray*}
\lefteqn{ p_{\left\langle \mathbf{Y}\right\rangle | \left\langle \mathbf{X}\right\rangle} (V|U)  } \\ & = & \sum_{r = 0}^h p_{\left\langle \mathbf{Y}\right\rangle | \left\langle \mathbf{X}\right\rangle, {\overline{\tau}}_{\mathbf{G}}} (V|U,r) \cdot p_{{\overline{\tau}}_{\mathbf{G}}}(r)  \\
 & = & p_{\left\langle \mathbf{Y}\right\rangle | \left\langle \mathbf{X}\right\rangle, {\overline{\tau}}_{\mathbf{G}}} (V|U,h-\dim V) \cdot  p_{{\overline{\tau}}_{\mathbf{G}}}(h-\dim V),
 \end{eqnarray*}
 which completes the proof.

 \end{proof}

From Theorem \ref{th_1}, we should note that the conditional probability $p_{\left\langle \mathbf{Y}\right\rangle | \left\langle \mathbf{X}\right\rangle} (V|U)$ depends on the dimension of $V$ (but not on $V$ itself) and is independent of $U$ given that $V \subseteq U$. Since the number of subspaces of dimension $\dim V$ of a subspace~$U$ of dimension $h$ is independent of the specific subspace~$U$, we can clearly see that our DMC is uniformly dispersive. However, it is not strongly symmetric in general. To see this, consider the conditional probability $p_{\left\langle \mathbf{Y}\right\rangle | \left\langle \mathbf{X}\right\rangle} (U|U) = p_{{\overline{\tau}}_{\mathbf{G}}}(0)$. The output subspace $U$ of dimension $h$ is reached by only one input subspace, namely, $U$ itself. Consider now the conditional probability $p_{\left\langle \mathbf{Y}\right\rangle | \left\langle \mathbf{X}\right\rangle} (O|U) = p_{{\overline{\tau}}_{\mathbf{G}}}(h)$. The zero-dimensional output subspace $O$ is reached by all input subspaces. From these two situations, we can clearly see that our DMC is not uniformly focusing, so it is not strongly symmetric in general. Nevertheless, we prove next an important result.
\lemma \label{le_1} The DMC $( {\cal P}(\mathrm{I\!F}_q^T,h), p_{\left\langle \mathbf{Y}\right\rangle | \left\langle \mathbf{X}\right\rangle}, \bigcup_{r=0}^h {\cal P}(\mathrm{I\!F}_q^T,r) )$ is symmetric.

\begin{proof}
There are $h+1$ component channels, DMC${}_r = ({\cal X}_r, p_r(y|x), {\cal Y}_r)$, $r=0, \ldots, h$, each one related to one of the possible rank deficiencies of the transfer matrix $\mathbf{G}$. As explained above, for ${\overline{\tau}}_{\mathbf{G}} = 0$ and for ${\overline{\tau}}_{\mathbf{G}} = h$, the two component channels are DMC${}_0 = ({\cal P}(\mathrm{I\!F}_q^T,h), p_0(y|x)=1 \mbox{ if } x=y, {\cal P}(\mathrm{I\!F}_q^T,h))$ and DMC${}_h = ({\cal P}(\mathrm{I\!F}_q^T,h),p_h(y|x)=1,\{ O \})$, respectively. These channels are trivial and clearly strongly symmetric. Their capacities are
\[ C_0 = \log {\cal P}(\mathrm{I\!F}_q^T,h) = \log  \dbinom{T}{h}_q, \]
and $C_h=0$.
Let us then assume that ${\overline{\tau}}_{\mathbf{G}} = \rho$, where $0 < \rho < h$. The input alphabet is ${\cal X}_{\rho} = {\cal P}(\mathrm{I\!F}_q^T,h)$. In the bipartite graph representation of DMC${}_{\rho}$, only those sockets of the ${\cal X}$ side of DMC that are associated to the transitions to output subspaces of dimension $h - \rho$ are considered for the ${\cal X}_{\rho}$ side. From (\ref{eq_t1b}), we can see that the output alphabet 
is ${\cal Y}_{\rho} =  {\cal P}(\mathrm{I\!F}_q^T,h-\rho) $, and every input (resp., output) subspace $U \in {\cal X}_{\rho}$ (resp., $V \in {\cal Y}_{\rho}$) is connected with equal transition probability to the same number of output (resp., input) subspaces.
(The number of input subspaces connected to an output subspace $V \in {\cal Y}_{\rho}$ can be obtained by dividing the number of sockets on the ${\cal Y}_{\rho}$ side, which is the same as the number of sockets on the ${\cal X}_{\rho}$ side,~{\it i.e.},
\[ {|{\cal P}(\mathrm{I\!F}_q^T,h)| \cdot \dbinom{h}{h- \rho}_q }, \]
by the number of output subspaces, namely, $|{\cal P}(\mathrm{I\!F}_q^T,h-\rho)|$.)
This yields a component strongly symmetric channel, with capacity given by (from Theorem \ref{th_ss})
\begin{eqnarray}
\label{eq_Cr}
C_{\rho} & = & \log |{\cal P}(\mathrm{I\!F}_q^T,h-\rho)| - \log |{\cal P}(\mathrm{I\!F}_q^h,h-\rho)| \nonumber \\
  & = & \log \dbinom{T}{h- \rho}_q - \log \dbinom{h}{h- \rho}_q \nonumber \\
  & = & \log \dfrac{\dbinom{T}{h- \rho}_q}{\dbinom{h}{h- \rho}_q}.
  \end{eqnarray}
\end{proof}

\theorem
\label{th_2} Given its rank deficiency probability distribution $p_{{\overline{\tau}}_G}(r)$, a RLCN seen as a subspace channel can be modeled as the DMC $ ( {\cal P}(\mathrm{I\!F}_q^T,h), p_{\left\langle \mathbf{Y}\right\rangle | \left\langle \mathbf{X}\right\rangle}, \bigcup_{r=0}^h {\cal P}(\mathrm{I\!F}_q^T,r) ). $ The capacity of this DMC is conditionally independent of the network topology and the network code used, and is given by
\[ \displaystyle C =\sum_{r=0}^h p_{{\overline{\tau}}_G}(r) \log \frac{\dbinom{T}{h-r}_{q}}{\dbinom{h}{h-r}_{q}}.
\]
\begin{proof}
The proof follows straightforwardly from Lemma~\ref{le_1} and Theorem \ref{th_s}, with $L=h+1$, $q_i = p_{{\overline{\tau}}_G}(i)$, and $C_i$ given by (\ref{eq_Cr}).
\end{proof}

\section{Conclusion and Final Comments}
\label{sec4}

In this paper, we have modeled random linear coding networks as a symmetric discrete memoryless channel, and we have derived the capacity of this channel. We have assumed that neither the transmitter nor the receiver has knowledge of the network transfer matrix (noncoherent scenario) or the network topology. This scenario differs from the one in~\cite{Yang.09}, where the network topology is assumed known and fixed at all times. For some fixed topologies, it has been shown in~\cite{Yang.09} that the capacity of RLCNs is lower bounded by the capacity of the so-called S-networks, for which constant-dimension subspace coding is optimal.

Herein, we have tactically considered the more general situation as an attempt to complement their work, and to try to answer some of the open questions raised therein. From the results of the present paper, we have the following comments:
\begin{enumerate}
\item The capacity in~\cite{Yang.09} is also a lower bound for the capacity of noncoherent RLCNs with unknown topology; if constant-dimension subspace coding is used, this lower bound is achieved;
\item The capacity of noncoherent RLCNs, either with general, but fixed and known, or with unknown topology, and whether subspace coding (in contrast to matrix coding) is sufficient to achieve capacity is still unkown.
\end{enumerate}

As with any DMC, the capacity of RLCNs is achieved by using the channel many times. Since subspace codes are the most promising candidates for noncoherent RLNC, {\em multishot} subspace codes have been investigated in~\cite{Roberto.09,Roberto.10}.

Finally, the case of non-constant-dimension subspace coding and the inclusion of erroneous packets to our channel model are currently being investigated.

\section*{Acknowledgements}
The authors would like to thank CNPq (Brazil) for the financial support, and Dr. Danilo Silva for helpful discussions. The first author would like to thank Dr. Yonghui Li, Dr. Zihuai Lin, and Prof. Branka Vucetic of the Telecommunications Laboratory of the University of Sydney, Australia, where this work was performed.

\bibliography{capacity_RLCN_ISIT2010}

\begin{thebibliography}{10}
\providecommand{\url}[1]{#1}
\csname url@samestyle\endcsname
\providecommand{\newblock}{\relax}
\providecommand{\bibinfo}[2]{#2}
\providecommand{\BIBentrySTDinterwordspacing}{\spaceskip=0pt\relax}
\providecommand{\BIBentryALTinterwordstretchfactor}{4}
\providecommand{\BIBentryALTinterwordspacing}{\spaceskip=\fontdimen2\font plus
\BIBentryALTinterwordstretchfactor\fontdimen3\font minus
  \fontdimen4\font\relax}
\providecommand{\BIBforeignlanguage}[2]{{%
\expandafter\ifx\csname l@#1\endcsname\relax
\typeout{** WARNING: IEEEtran.bst: No hyphenation pattern has been}%
\typeout{** loaded for the language `#1'. Using the pattern for}%
\typeout{** the default language instead.}%
\else
\language=\csname l@#1\endcsname
\fi
#2}}
\providecommand{\BIBdecl}{\relax}
\BIBdecl

\bibitem{ho.03}
T.~Ho, R.~Koetter, M.~Médard, D.~Karger, and M.~Effros, ``The benefits of
  coding over routing in a randomized setting,'' in \emph{Proc. 2003 IEEE Int.
  Symp. Information Theory (ISIT'03)}, Yokohama, Japan, 2003, p. 442.

\bibitem{Ahlswede.00}
R.~Ahlswede, N.~Cai, S.-Y.~R. Li, and R.~W. Yeung, ``Network information
  flow,'' \emph{IEEE Trans. Inform. Theory}, vol.~46, no.~4, pp. 1204--1216,
  Jul. 2000.

\bibitem{LNC.03}
S.-Y.~R. Li, R.~W. Yeung, and N.~Cai, ``Linear network coding,'' \emph{IEEE
  Trans. Inform. Theory}, vol.~49, no.~2, pp. 371--381, Feb. 2003.

\bibitem{Yang.09}
S.~Yang and E.-H. Yang, ``The worst network topology for random linear network
  coding,'' unpublished preprint, 2009.

\bibitem{Silvacapacity.08}
D.~Silva, F.~Kschischang, and R.~Koetter, ``Communication over finite-field
  matrix channels,'' \emph{Computing Research Repository (CoRR), submitted to
  IEEE Trans. Inform. Theory}, vol. abs/0807.1372, Jul. 2008.

\bibitem{Monta.07}
A.~Montanari and R.~Urbanke, ``Coding for network coding,'' \emph{Computing
  Research Repository (CoRR)}, vol. abs/0711.3935, Nov. 2007.

\bibitem{Silva.08}
D.~Silva, F.~Kschischang, and R.~Koetter, ``A rank-metric approach to error
  control in random network coding,'' \emph{IEEE Trans. Inform. Theory},
  vol.~54, no.~9, pp. 3951--3967, Sep. 2008.

\bibitem{Jafari.08}
M.~J. Siavoshani, C.~Fragouli, and S.~Diggavi, ``Noncoherence multisource
  network coding,'' in \emph{Proc. 2008 IEEE Int. Symp. Information Theory
  (ISIT'08)}, Toronto, Canada, Jul. 2008, pp. 817--821.

\bibitem{Jafari.09}
C.~Fragouli, M.~Jafari, S.~Mohajer, and S.~Diggavi, ``On the capacity of
  non-coherent network coding,'' in \emph{Proc. 2009 IEEE Int. Symp.
  Information Theory (ISIT'09)}, Seoul, Korea, 2009, pp. 273--277.

\bibitem{KS.08}
R.~Koetter and F.~R. Kschischang, ``Coding for errors and erasures in random
  network coding,'' \emph{IEEE Trans. Inform. Theory}, vol.~54, no.~8, pp.
  3579--3591, Aug. 2008.

\bibitem{Gabidulin.85}
E.~M. Gabidulin, ``Theory of codes with maximum rank distance,'' \emph{Probl.
  Inf. Transm.}, vol.~21, no.~1, pp. 1--12, 1985.

\bibitem{Silvaisit.09}
D.~Silva and F.~Kschischang, ``Fast encoding and decoding of {G}abidulin
  codes,'' in \emph{Proc. 2009 IEEE Int. Symp. Information Theory (ISIT'09)},
  Seoul, Korea, 2009, pp. 2858--2862.

\bibitem{Cover.06}
T.~Cover and J.~Thomas, \emph{Elements of Information Theory}.\hskip 1em plus
  0.5em minus 0.4em\relax Second Edition, New York: John Wiley \& Sons, 2006.

\bibitem{Massey.93}
J.~L. Massey, \emph{Lecture Notes for Applied Digital Information Theory
  I}.\hskip 1em plus 0.5em minus 0.4em\relax Abteilung für Elektrotechnik: ETH,
  Zürich, 1993.

\bibitem{Roberto.09}
R.~W. da~N\'{o}brega and B.~F. Uch\^{o}a-Filho, ``Multishot codes for network
  coding: bounds and a multilevel construction,'' in \emph{Proc. 2009 IEEE Int.
  Symp. Information Theory (ISIT'09)}, Seoul, Korea, 2009, pp. 428--432.

\bibitem{Roberto.10}
------, ``Multishot codes for network coding using rank-metric codes,''
  submitted to {\it 2010 IEEE Int. Symp. Information Theory (ISIT'10)}.

\end{thebibliography}
\bibliographystyle{ieeetran}

\end{document}